\documentclass[aps,prc,superscriptaddress,showpacs,manuscript]{revtex4}
\usepackage{amsmath,bm}%
\usepackage{graphicx}%
\begin{document}
\title{Isocaling and the Symmetry Energy in the Multifragmentation Regime of Heavy Ion Collisions}
\author{Z. Chen}
\affiliation{Cyclotron Institute, Texas A$\&$M University, College Station, Texas 77843}
\affiliation{Institute of Modern Physics, Chinese Academy of Sciences, Lanzhou, 730000,China.}
\author{S. Kowalski}
\affiliation{Institute of Physics, Silesia University, Katowice, Poland.}
\author{M. Huang}
\affiliation{Cyclotron Institute, Texas A$\&$M University, College Station, Texas 77843}
\affiliation{Institute of Modern Physics, Chinese Academy of Sciences, Lanzhou, 730000,China.}
\affiliation{Graduate University of Chinese Academy of Sciences, Beijing, 100049, China.}
\author{R. Wada}
\email[E-mail at:]{wada@comp.tamu.edu}
\affiliation{Cyclotron Institute, Texas A$\&$M University, College Station, Texas 77843}
\author{T. Keutgen}
\affiliation{FNRS and IPN, Universit\'e Catholique de Louvain, B-1348 Louvain-Neuve, Belgium.}
\author{K. Hagel}
\affiliation{Cyclotron Institute, Texas A$\&$M University, College Station, Texas 77843}
\author{J. Wang}
\affiliation{Institute of Modern Physics, Chinese Academy of Sciences, Lanzhou, 730000,China.}
\author{L. Qin}
\affiliation{Cyclotron Institute, Texas A$\&$M University, College Station, Texas 77843}
\author{J.B. Natowitz}
\affiliation{Cyclotron Institute, Texas A$\&$M University, College Station, Texas 77843}
\author{T. Materna}
\affiliation{Cyclotron Institute, Texas A$\&$M University, College Station, Texas 77843}
\author{P.K. Sahu}
\affiliation{Cyclotron Institute, Texas A$\&$M University, College Station, Texas 77843}
\author{A. Bonasera}
\affiliation{Cyclotron Institute, Texas A$\&$M University, College Station, Texas 77843}
\affiliation{Laboratori Nazionali del Sud, INFN,via Santa Sofia, 62, 95123 Catania, Italy}
%\affiliation{Libera Universita Kore di Enna, 94100 Enna, Italy}

%\author{Z. Chen, S. Kowalski, M. Huang, R. Wada, T. Keutgen, K. Hagel, J. Wang,
%          L. Qin, J.B. Natowitz, T. Materna, P.K. Sahu, and A. Bonasera }

\date{\today}
 
\begin{abstract}
The ratio of the symmetry energy coefficient to temperature, 
$a_{sym}/T$, in Fermi energy heavy ion collisions, has been experimentally
extracted as a function of the fragment atomic number using isoscaling parameters and 
the variance of the isotope distributions. The extracted values have
been compared to the results of  calculations made with  an 
Antisymmetrized Molecular Dynamics
(AMD) model employing  a statistical decay code to account for 
deexcitation of excited primary fragments. The experimental
values are in good agreement with the values calculated but are  
significantly different from those characterizing the yields of 
the primary AMD fragments.
  
\end{abstract}
\pacs{21.65.Ef, 24.10.-i, 24.10.Pa,25.70.Gh, 25.70.Pq}
\maketitle

\section *{I. INTRODUCTION}

In Fermi energy heavy ion collisions, fragments are copiously produced. 
The mass distributions of these fragments exhibit a power low behavior 
which has been
discussed in terms of the Modified Fisher Model~\cite{Minich82,Bonasera08}.
The isotope distributions of these fragments play a key role in these
analyses. Theoretical studies indicate that the isotope formation is 
governed by the
free energy at the density and temperature of the emitting system. 
The experimental observation of  isoscaling for two similar
reactions with different neutron to proton ratios, N/Z, demonstrates
that the free energies and therefore the yields of the fragments are 
also closely related to the N/Z of the emitting
system~\cite{Xu00,Tsang01,Botvina02,Ono03}. Thus the experimental yield 
of isotope with N neutrons and Z protons
can be given by~\cite{Minich82,Bonasera08,Albergo85,Tsang01}:
\begin{eqnarray}
Y(N,Z)=Y_0F(N,Z)A^{-\tau}exp\{-[G(N,Z) \nonumber \\
-\mu_nN-\mu_pZ]/T\}
\label{eq:yield}
\end{eqnarray}
where $Y_{0}$ is a constant and G(N,Z) is the nuclear free energy 
at the time of the fragment formation. 
$\mu_{n}$ and $\mu_{p}$ are the chemical potential of neutron
and proton, and T is the temperature of the emitting source. The factor 
F(N,Z) is the correction factor
for the feeding from the statistical decay processes. The factor, 
$A^{-\tau}$, originates from the entropy of the fragment~\cite{Minich82}.
The symmetry energy term in the free energy, G(N,Z), is usually expressed as:
\begin{equation}
E_{sym}=a_{sym}(N-Z)^{2}/A
\label{eq:symmetry}
\end{equation}
where A=N+Z and $a_{sym}$ is the symmetry energy coefficient which  depends
on the  nuclear density $\rho$ and the temperature T of the emitting 
source.

From Eq.(\ref{eq:yield}) $R_{12}$, the ratio of the isotope yields for
two similar reaction systems with different N/Z ratios, can be written as:
\begin{equation}
R_{12}(N,Z)=Cexp(\alpha N + \beta Z)
\label{eq:isoscaling}
\end{equation}
This relation is known as the isoscaling relation. The isoscaling 
parameters,  $\alpha = (\mu_{n}^{1} - \mu_{n}^{2})/T$
and $\beta = (\mu_{p}^1 - \mu_{p}^2)/T$) are the differences  of the neutron
or proton chemical potentials between the systems 1 and 2, divided by the
temperature. C is a constant. System 1 is normally taken as the more 
neutron rich of the two.

As discussed in refs.~\cite{Botvina02,Moretto08,Ono04_1}, the isoscaling 
parameters and the symmetry
energy coefficient are closely related. For a multifragmentation regime,
as pointed out in ref.~\cite{Ono04_1}, this relation is given by:
\begin{equation}
\alpha(Z) = 4a_{sym}\Delta (Z/\bar{A})^{2}/T
\label{eq:alpha}
\end{equation}
where $\Delta(Z/\bar{A})^2 = {(Z/\bar{A})^{2}}_{1} - {(Z/\bar{A})^{2}}_{2}$
for the two reaction systems and $\bar{A}$ is the average mass number of 
isotopes for a given Z.

There are two issues for the determination of the $a_{sym}$ values in
Eq.(\ref{eq:alpha}). One is the source temperature T.  Since the beginning
of the experimental study of heavy ion collisions in the
multifragmentation regime, significant efforts have been made to extract 
the source temperature, but different
methods of temperature extraction can lead to different results and 
uncertainties still remain~\cite{Kelic06}. Another
issue is the effect of the secondary decay process, 
expressed by F(N,Z) in Eq.(\ref{eq:yield}). In experiments fragments
have typically cooled down to the ground state before they are detected.
Indeed, in previous works, excitation energies of the primary fragments 
have been evaluated by studying the associated light charged particle 
multiplicities~\cite{Marie98,Hudan03}. Such data raise the question of 
the degree of confidence for the experimentally extracted symmetry energy 
coefficient, in which this important effect is not properly corrected. 
In fact in a separate paper using the same data set 
presented here we have demonstrated that the secondary 
decay processes significantly effect the isobaric yield ratios and the 
experimentally extracted symmetry energy 
coefficient~\cite{Huang10}. 
In this paper, we focus on the relation between the ratio $a_{sym}/T$ 
and isoscaling parameters and between the ratio and the widths of 
the isotope distributions. The experimentally extracted values 
of $a_{sym}/T$ extracted from both observables are compared to those extracted from the model calculations.

\section *{II. EXPERIMENT}

The experiment was performed at the K-500 superconducting cyclotron 
facility at Texas A$\&$M University. $^{64,70}$Zn and $^{64}$Ni beams 
were used to irradiate $^{58,64}$Ni, $^{112,124}$Sn, $^{197}$Au and 
$^{232}$Th targets at 40 A MeV. Intermediate mass fragments (IMFs) were 
detected by a detector telescope placed at 20$^\circ$. The telescope 
consisted of four Si detectors. Each Si detector was 5cm x 5cm. The 
nominal thicknesses were 129, 300, 1000, 1000 $\mu$m. All Si detectors 
were segmented into four sections and each quadrant had a 5$^\circ$ 
opening angle in polar and azimuthal angles. Therefore the energies 
of the fragments 
were measured at two polar angles of the quadrant detector, namely $\theta$ 
= 17.5$^\circ$ $\pm$ 2.5$^\circ$ and $\theta$ = 22.5$^\circ$ $\pm$ 
2.5$^\circ$. Typically 6-8 isotopes for a given atomic numbe up to Z=18 
were clearly identified with the energy threshold of 4-10 A MeV, 
using the $\Delta$E-E technique for any two consecutive detectors. 
The $\Delta$E-E spectrum was linearized  empirically. 
Mass identification of the isotopes 
were made using a range-energy table~\cite{Hubert90}. In the analysis 
code, isotopes are identified by a parameter $Z_{Real}$. For the isotope 
with A=2Z, $Z_{Real}$ = Z is assigned and other isotopes are identified 
by interpolation between them. Typical $Z_{Real}$ spectra are shown 
in Fig.1. The energy spectrum of each isotope was extracted by gating 
the isotope in a 2D plot of $Z_{Real}$ vs energy. The yields of light 
charged particles (LCPs) in coincidence with IMFs were also measured 
using 16 single crystal CsI(Tl) detectors of 3cm thickness set around 
the target. The light output from each detector was read by a photo 
multiplier tube. The pulse shape discrimination method was used to 
identify p, d, t, h and $\alpha$ particles. The energy calibration 
for these particles were performed using Si detectors (50 -300 $\mu$m) 
in front of the CsI detectors in separate runs.

The yield of each isotope was evaluated, using a moving source fit. 
For LCPs, three sources (projectile-like(PLF), nucleon-nucleon-like(NN)
and target-like (TLF)) were used. The NN-like sources have source 
velocities of about a half of the beam velocity. The parameters are 
searched globally for all 16 angles. For IMFs, since the energy spectra 
were measured only at the two angles of the quadrant detector, the 
spectra were parameterized using a single NN-source. Using a source 
with a smeared source velocity around half the beam velocity, the 
fitting parameters were first determined from the spectrum summed 
over all isotopes for a given Z, assuming A=2Z. Then all extracted 
parameters except for the normalizing yield parameter were used for 
the individual isotopes. This procedure was based on the assumption 
that, when the spectrum is plotted in energy per nucleon, the shape 
of the energy spectrum is the same for all isotopes for a given Z. 
Indeed the observed energy spectra of isotopes are well reproduced by this 
method. For IMFs, a further correction was made for the background. 
As seen in Fig.1, the isotopes away from the 
stability line, such as $^{10}$C and $^{29}$Mg, have a very small yields 
and the background contribution is significant. In order to evaluate 
the background contribution to the extracted yield from the source fit, 
a two Gaussian fit to each isotope combined with a linear background 
was used. The fits are shown in Fig.1. Each peak consists of two 
Gaussians. The second Gaussian (about 10\% of the height of the first one)
is added to reproduce the shape of the valley between two isotopes. 
This component is attributed to the reactions of the isotope in the Si 
detector. The centroid of the Gaussians was set to the value calculated 
from the range-energy table within a small margin. The final yield of 
an isotope was determined  by correcting the yield evaluated 
from the moving source fit by the ratio between the two Gaussian yields 
and the linear background. Rather large errors ( $\sim \pm10\%$) are 
assigned for the multiplicity of the NN source for IMFs, originating 
from the source fit besides the background estimation. The errors from the 
source fit are evaluated from the different assumptions 
of the parameter set for the source velocity and temperature.

\begin{figure}[ht]
\includegraphics[width=3.6in]{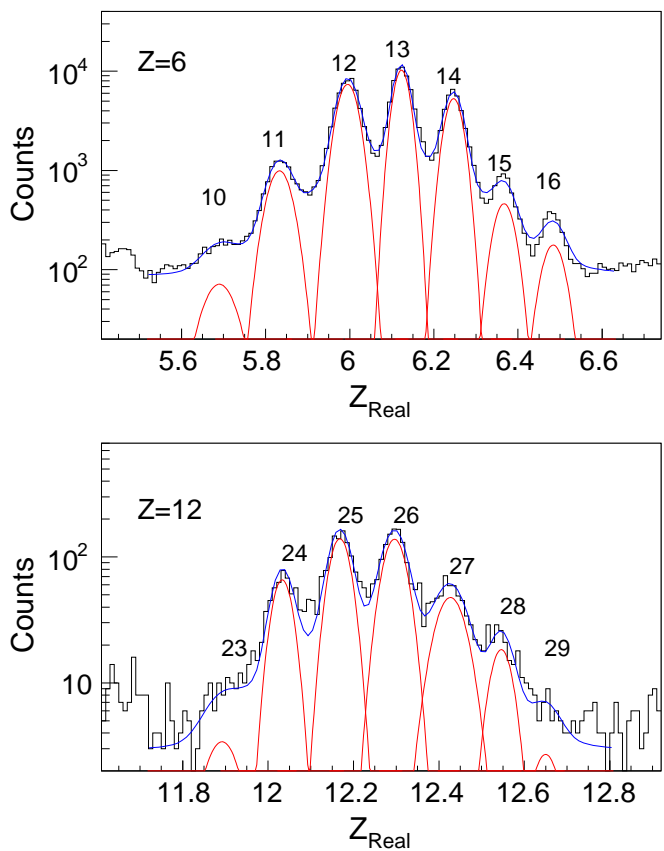}
\caption{\label{fig:fig_1} Typical linearized isotope spectra are shown 
for Z=6 (upper figure)  and 12 (lower)
cases. The number at the top of each peak is the mass number assigned.
Linear back ground is assumed from valley to valley for a given Z.
Each Gaussian indicates the yield of the isotope above the back ground.}
\end{figure}

\section *{III. ISOSCALING}

\begin{figure}[ht]
\includegraphics[width=3.6in]{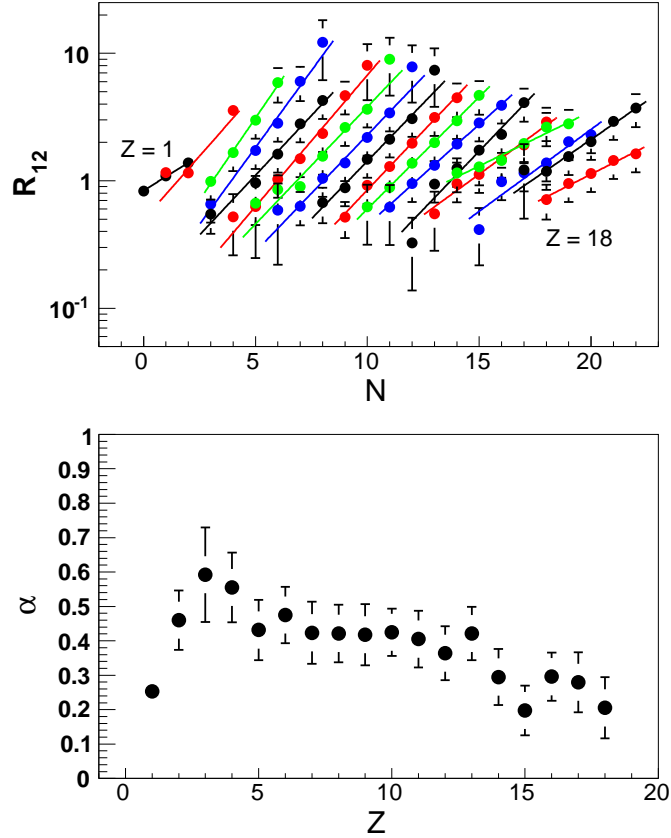}
\caption{\label{fig:fig_2}(upper) $\alpha(Z)$ values as a function of Z for 
$^{64}$Ni + $^{124}$Sn 
and $^{64}$Ni + $^{58}$Ni. From the left to right each lines 
correspond to Z=1 to Z=18. (lower) The extracted $\alpha(Z)$ values are 
plotted as a function of Z.}
\end{figure}

At the top of Fig.~\ref{fig:fig_2}, the yield ratio of 
Eq.(\ref{eq:isoscaling}) for the reactions $^{64}$Ni + $^{124}$Sn 
and $^{64}$Ni + $^{58}$Ni is plotted as a 
function of N. The $\alpha$ parameter is determined by 
individual fits to yield ratios for isotopes with 
a given Z. The extracted values are plotted in the bottom of 
Fig.~\ref{fig:fig_2}. As seen in the figure, the extracted $\alpha(Z)$ parameter shows a steady decrease as Z increases for Z $\ge$ 4. The $\beta(N)$ parameter generally shows much similar variation with increasing N, and has the opposite sign. Hereafter $\alpha$ and $\beta$ are denoted as 
$\alpha(Z)$ and $\beta(N)$. Isoscaling parameters, $\alpha(Z)$ and $\beta(Z)$,
have been evaluated for all possible combinations between two reactions. 
For 18 different reactions are considered here. More than 150 different 
combinations have been studied. 

\begin{figure}[ht]
\includegraphics[width=3.6in]{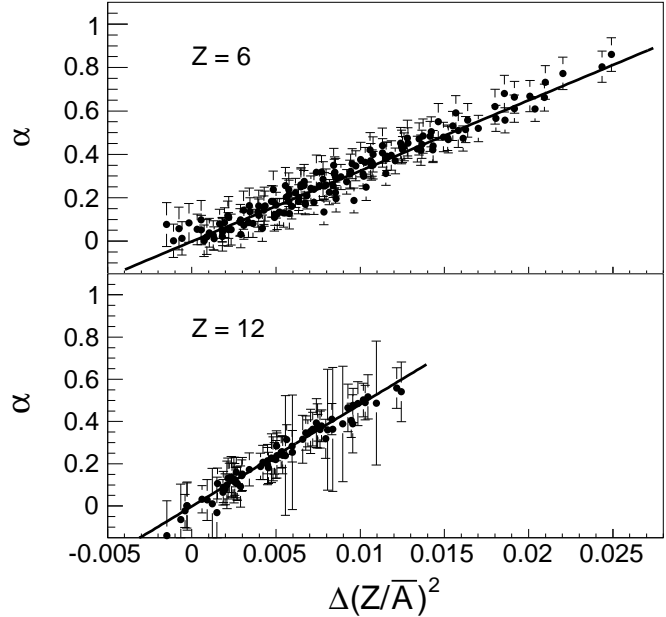}
\caption{\label{fig:fig_4}$\alpha(Z)$ values as a function of  
$\Delta(Z/A)^{2}$ for Z=6 (upper) and Z=12 (lower).
 The lines are the results of a linear fit.}
\end{figure}

In Fig.~\ref{fig:fig_4} the extracted $\alpha(Z)$ values are plotted 
as a function of $\Delta(Z/\bar{A})^{2}$ for Z=6 and Z=12.  Each data 
point represents a combination of two reactions. As seen in the figure, 
the $\alpha(Z)$ values are linearly related to $\Delta(Z/\bar{A})^{2}$. 
The slope associated with this relationship increases gradually as Z 
increases. The correlations have been fit by a linear function for 
each Z and the slope values, which correspond to the value 4$a_{sym}$/T 
in Eq.(\ref{eq:alpha}), have been extracted. 

\begin{figure}[ht]
\includegraphics[width=3.6in]{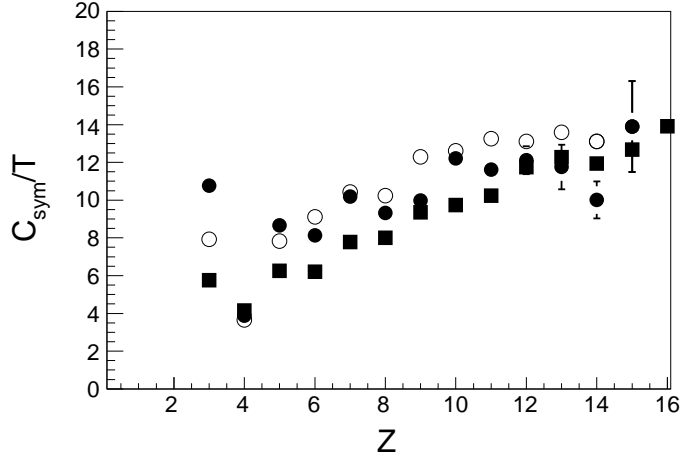}
\caption{\label{fig:fig_5} Experimental $a_{sym}(\rho,T)/T$ values 
extracted from Eq.(\ref{eq:alpha}) are shown by solid circles as a
function of Z. Open circles show calculated $a_{sym}(\rho,T)/T$ values for 
the fragments yields in AMD-Gemini calculations. 
Squares are experimental results of $\zeta(Z)$ discussed in the 
section IV.}
\end{figure}

In Fig.~\ref{fig:fig_5} the extracted values of  $a_{sym}$/T are 
plotted as a function of Z and shown by solid circles. A clear trend is 
observed for the 
parameter, $a_{sym}$/T. The value increases from 4 to 14 as Z 
increases from 4 to 15. The value for Z=3 is much larger than Z=4. 
This is partially caused by the isotope distribution. Since $^{5}$Li 
is unstable and decays  before arriving to the detector, $\bar{A}$  
deviates from the actual centroid of the isotope distribution. 
An attempt has been made to fit the distribution by a Gaussian 
function and determine $\bar{A}$ as the centroid value. This procedure 
makes $a_{sym}$/T around 3 for Z=3, but the uncertainty is significant, 
especially for neutron deficient systems. Therefore in the plot the 
experimental Z/$\bar{A}$ value appears without correction. 

\section *{IV. SYMMETRY ENERGY AND VARIANCE OF THE ISOTOPE DISTRIBUTION}

\begin{figure}[ht]
\includegraphics[width=3.6in,clip]{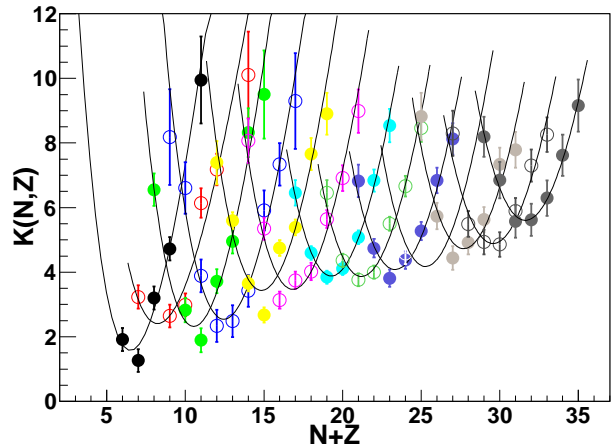}
\caption{\label{fig:fig_6} K(N,Z) distribution from five reaction 
systems, 
$^{64}$Ni+$^{58}$Ni, $^{64}$Ni,$^{112}$Sn, $^{197}$Au, $^{232}$Th.}
\end{figure}

The multiplicity distributions of the isotopes for a given Z show a 
quadratic distribution when they
are plotted on a logarithmic scale. Since the symmetry energy term is 
the only term  proportional to (N-Z)$^{2}$ in the free energy,  
this suggests that the variance of the distributions is closely related 
to the symmetry energy
coefficient. In order to explore the relation between the symmetry energy
term in the free energy and the variance of the isotope distribution,
Ono et al. introduced a generalized function K(N,Z) for the free energy
in  ref.~\cite{Ono04_1} as given below.
\begin{equation}
K(N,Z)=\sum_{i=1}^{n}w_i(N,Z)[-lnY_i(N,Z)+\alpha_i(Z)N+\gamma_i(Z)]
\label{eq:knz1}
\end{equation}
Here i represents each reaction. The summation is taken over i for the 
different N/Z reaction systems in order to get isotope multiplicity 
distribution in a wide range from proton rich to neutron rich isotopes. 
The average weights, 
w$_i$(N,Z), are determined by minimizing the statistical errors 
in K(N,Z) for a given (N,Z). The isoscaling parameter, $\alpha_i(Z)$, 
is the isoscaling parameter value evaluated in the previous section. 
For each Z the 
parameters, $\gamma_i$(Z), are determined by optimizing the agreement 
of the quantities [-lnY(N,Z) + $\alpha_i(Z)$ + $\gamma_i$(Z)] from 
different reactions. A typical
K(N,Z) distribution from the experiment is shown in 
Fig.~\ref{fig:fig_6}. 
The isotope distributions for a given Z exhibit a smooth quadratic 
distribution and they can be  fit by a function:
\begin{equation}
K(N,Z)=\xi(Z)N+\eta(Z)+\zeta(Z)(N-Z)^{2}/A
\label{eq:knz2}
\end{equation}
Where $\xi(Z)$, $\eta$(Z), $\zeta$(Z) are the fitting parameters. 
As one can see the functional form, $\zeta$(Z) is related to 
the symmetry energy coefficient given in Eq.(\ref{eq:alpha}) as:
\begin{equation}
\zeta(Z)=a_{sym}/T
\label{eq:zeta}
\end{equation}
In Fig.~\ref{fig:fig_5}, the values of $\zeta$(Z) extracted using this 
technique  are shown by solid squares. The values 
are generally about 1 or 2 units smaller than the $a_{sym}$ /T values 
evaluated in the previous section (solid circles), 
but the general trend is in good agreement. The difference in the 
extracted values for Z=3 originate from the different determination of the
average value of A. In the analysis in this section the average masses of 
these isotopes are determined by the centroid of the quadratic distributions.

\section *{V. COMPARISONS WITH MODEL SIMULATIONS}

\begin{figure} [ht]
\includegraphics[width=3.6in]{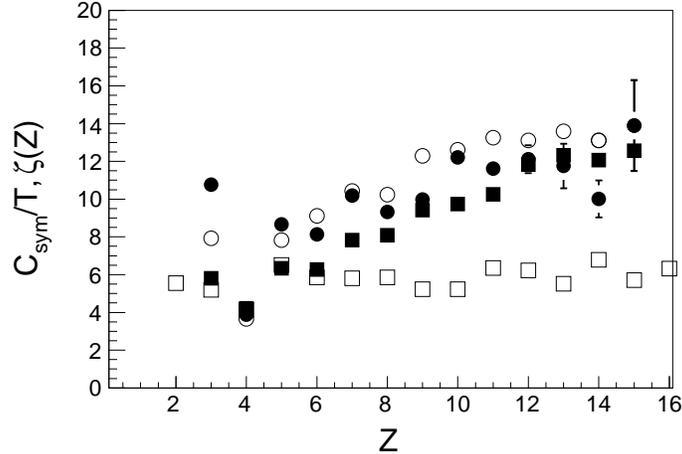}
\caption{\label{fig:fig_7} Comparisons between the experimentally 
extracted symmetry  energy 
coefficient to temperature ratio and those from AMD.  Solid circles show the 
experimental values determined from  Eq.(\ref{eq:alpha}) and 
open circles show results from of AMD-Gemini calculations. Closed 
squares indicate results for 
the primary fragments with Rc=5.}
\end{figure}

The experimentally detected fragments are the final products of the 
reaction. 
Excited primary fragments will have cooled down by statistical decay  
before they arrive in the detector. Excitation energies of the
primary fragments have been evaluated experimentally by measuring the 
light charged particle multiplicities in coincidence with the 
fragments~\cite{Marie98,Hudan03}. Typical excitation energies of 2 to 3 
MeV/nucleon have been derived. In order to study the effect of the 
secondary decay process on the experimentally extracted ratio, 
$a_{sym}$/T, the simulation codes of an Antisymmetrized Molecular 
Dynamics (AMD) model~\cite{Ono96,Ono99, Hudan06} and a statistical 
decay code, Gemini~\cite{Charity88}, have been 
used. These codes have often been used to study the 
fragment production in Fermi energy heavy ion reactions and the global 
features of the experimental results have been well
reproduced~\cite{Ono02,Wada98,Ono04_2,Wada00,Wada04,Hudan06}. Since 
the AMD calculation requires a lot of CPU time, only two of the 
experimental reaction systems have been studied. The systems examined 
are $^{64}$Zn+$^{112}$Sn and $^{64}$Ni+$^{124}$Sn at 40 AMeV. All 
calculations shown in this paper have been performed in a newly 
installed computer cluster 
in the Cyclotron Institute~\cite{Wada09}. In order to obtain yields of  
the final products, the deexcitation of primary 
fragments formed  at 300 fm/c was followed using the Gemini code until 
they cooled to the ground state. Using the same analysis described
in the section III, the scaling parameters, $\alpha(Z)$ and $\beta(Z)$ were 
then extracted from the simulated events as a 
function of Z. The average mass number of the isotopes for a given Z was 
also evaluated from the calculations. The calculated variation of  
$a_{sym}$/T is shown  by open circles 
in Fig.~\ref{fig:fig_5}. The values are typically one to two units 
higher than the experimental values (open circles) but exhibit an 
essentially identical trend. For the 
primary fragments at t = 300 fm/c the same analysis has been made. 
These fragments are identified using a coalescence 
technique in phase space. The evaluated symmetry coefficients using 
$R_c$ = 5 are shown in Fig.~\ref{fig:fig_7}. $R_{c}$=5 corresponds to 
a radius of 5 fm in configuration space.  For the primary fragments 
the calculated values are close to 
the experimental values observed  for Z $\leq$6 but, in contrast to 
the experimental results, they remain more or less constant for all Z. 
The rather flat distribution of $a_{sym}$/T values over the entire 
range of Z is consistent with the picture of the origin of the primary 
fragments from a common emitting source with a given density and 
temperature. These comparisons indicate that the $a_{sym}$/T values 
are significantly modified by the secondary process for Z $>$ 4

\section *{VI. SUMMARY}

The symmetry energy coefficient to temperature ratio, $a_{sym}$/T , 
as a function of Z has been extracted from experimental data in two 
different ways, namely from isoscaling parameters and from the variance 
of the observed  isotope distributions. The results from the two 
techniques are in reasonable agreement.  Experimental values increase 
from $\sim$ 4 to 12 as Z increases from 4 to 15. The values and trends 
observed for the final fragments are well reproduced by the AMD plus 
Gemini model simulations. However these values are significantly 
different from the values extracted for the primary fragments, 
especially for Z $>$ 4, suggesting 
that the derivation of the observed ratio of  symmetry energy coefficient to 
temperature is significantly perturbed  by the secondary decay processes.

\begin{acknowledgments}
We thank the staff of the Texas A$\&$M Cyclotron facility for their 
support during the experiment. We thank L. Sobotka for letting us to 
use his spherical scattering chamber. We also thank A. Ono and R. 
Charity for letting us to use their calculation 
codes. This work is supported by the U.S. Department of Energy under 
Grant No. DE-FG03-93ER40773 and the Robert A. Welch Foundation under 
Grant A0330. One of us(Z. Chen) also thanks the \textquotedblleft100 
Persons Project" of the Chinese Academy of Sciences for the support.
\end{acknowledgments}


\begin{thebibliography}{}
\bibitem{Minich82}R. W. Minich {\it et al.}, Phys. Lett. {\bf B118}, 
458 (1982).
\bibitem{Bonasera08}A. Bonasera {\it et al.}, Phys. Rev. Lett. {\bf 101}, 
122702 (2008).
\bibitem{Xu00}H. S. Xu {\it et al.}, Phys. Rev. Lett. {\bf 85}, 4, (2000).
\bibitem{Tsang01}M. B. Tsang {\it et al.}, Phys. Rev. {\bf C64}, 054615 
(2001).
\bibitem{Botvina02}A.S. Botvina, O. V. Lozhkin and W. Trautmann, Phys. Rev.
{\bf C65}, 044610 (2002).
\bibitem{Ono03}A. Ono {\it et al.}, Phys. Rec. {\bf C68}, 051601(R) (2003).
\bibitem{Albergo85}S. Albergo {\it et al.}, Nuovo Cimento, {\bf A89}, 1 
(1985).
\bibitem{Moretto08}L. G. Moretto, C. O. Dorso, J. B. Elliott, and L. Phair, 
Phys. Rev. {\bf C77}, 037603 (2008).
\bibitem{Ono04_1}A. Ono, {\it et al.}, Phys. Rev. {\bf C70}, 041604(R) 
(2004).
\bibitem{Kelic06}A. Keli$\acute{c}$, J. B. Natowitz and K. H.Schmidt, Eur. 
Phys. J.{\bf A30},203 (2006).
\bibitem{Marie98}N. Marie {\it et al.}, Phys. Rev. {\bf C58}, 256 (1998).
\bibitem{Hudan03}S. Hudan {\it et al.}, Phys. Rev. {\bf C76}, 064613 (2003)
\bibitem{Huang10}M.Huang {\it et al.}, arXiv:1001.3621 [nucl-ex] 22Jan2010.
\bibitem{Wada05}R.Wada {\it et al.}, annual report of the Cyclotron 
Institute,
 Texas A\&M University, (2005), {\bf II-3}, unpublished.
\bibitem{Hubert90}F. Hubert, R. Bimbot and H. Gauvin, At. Data Nucl. Data
Tables {\bf 46}, 1 (1990).
\bibitem{Ono96}A. Ono and H. Horiuchi, Phys. Rev {\bf C53}, 2958 (1996).
\bibitem{Ono99}A. Ono, Phys Rev {\bf C59}, 853 (1999)
\bibitem{Ono04_2}A. Ono and H. Horiuchi, Prog. Part. Nucl. Phys. {\bf53}, 
501 (2004).
\bibitem{Charity88}R. J. Charity {\it et al.}, Nucl. Phys. {\bf A483}, 
371 (1988).
\bibitem{Ono02}A. Ono, {\it et al.}, Phys. Rev. {\bf C66}, 014603 (2002).
\bibitem{Wada98}R. Wada {\it et al.}, Phys. Lett. {\bf B422}, 6, (1998).
\bibitem{Wada00}R. Wada {\it et al.}, Phys. Rev. {\bf C62}, 034601 (2000).
\bibitem{Wada04}R. Wada {\it et al.}, Phys. Rev. {\bf C69}, 044610 (2004)
\bibitem{Hudan06}S. Hudan, R. T. de Souza and A. Ono, Phys. Rev. 
{\bf C73}, 054602 (2006).
\bibitem{Wada09}R.Wada {\it et al.}, annual report of the Cyclotron 
Institute,
Texas A\&M University, (2009), unpublished.

\end{thebibliography}
\end{document}